\pdfoutput=1
\documentclass[aps,prd,amsmath,floats,floatfix,superscriptaddress,nofootinbib,showpacs]{revtex4-2}

\usepackage[T1]{fontenc}
\usepackage[utf8]{inputenc}
\usepackage{lmodern}
\usepackage{lipsum}
\usepackage[dvipsnames, usenames]{xcolor}
\definecolor{linkcolor}{rgb}{0.0,0.3,0.5}
\usepackage[hypertexnames=false, unicode, colorlinks=true, linkcolor=linkcolor,
citecolor=linkcolor, filecolor=linkcolor,urlcolor=linkcolor,
pdfusetitle]{hyperref}
\usepackage[all]{hypcap}
\usepackage{graphicx}
\usepackage{xspace}
\usepackage{amssymb}
\usepackage[normalem]{ulem} 
\usepackage{bm} 
\usepackage{appendix}
\usepackage{microtype}
\usepackage[english]{babel}
\usepackage{blindtext}

\begin{document}

\title{Front Propagation from  
Radiative Sources}

\author{Theodore Steele}
\email{ts715@damtp.cam.ac.uk}
\affiliation{Department of Applied Mathematics and Theoretical Physics, University of Cambridge}

\author{Kinwah Wu}
\email{kinwah.wu@ucl.ac.uk}
\affiliation{Mullard Space Science Laboratory, University College London}

\begin{abstract}
Fronts are regions of transition from one state to another in a medium.  
They are present 
in many areas of science and applied mathematics, 
and modelling them and their evolution is often an effective way of treating the underlying phenomena responsible for them.  
In this paper, 
we propose a new 
approach to modelling front propagation,  
which characterises the evolution 
of structures  
surrounding radiative sources.  
This approach is generic  
  and has a wide range of applications, particularly when dealing with the propagation of 
  phase or state transitions in media 
  surrounding radiation emitting objects.  
As an illustration, 
  we show an application in modelling 
  the propagation 
  of ionisation fronts around early stars during the cosmological Epoch of Reionisation (EoR) and show that the results are consistent with those of existing equations but provide much richer sources of information. 
\end{abstract}

\maketitle

\section{Introduction}
A front is a transitional region between two states.  
They can be infinitesimal in thickness or may take 
the form of a wide gradient.  
They may also be either static or dynamic, 
 depending upon the nature 
 of the system 
 and states being modelled.  The study of the dynamics of fronts 
  is important to many areas 
  of sciences, in particular physics 
  and applied mathematics, 
  including the study of the effects of ionising radiation, of chemical catalysis by radiation, of the spread of fires, and of pressure shock waves.  

A number of algorithms 
   have been developed 
   to model the propagation of fronts 
   in various scenarios, many of which rely upon finding solutions to the Eikonal equation of motion. 
Examples of Eikonal based models 
  are the level-set method 
  and its variants 
  \cite{Sethian1996,GIBOU201882,doi:10.1137/0524066,Osher2003}, 
  which model the propagation of fronts 
  from a set of initial conditions. 
A simple yet useful class of algorithms 
  in the restricted level-set methods 
  are known as the fast-marching algorithms 
  \cite{doi:10.1073/pnas.93.4.1591,10.2307/2653069}, 
  in which 
  a monotonically evolving front 
  is defined only by its local conditions, again through treatment of the Eikonal equation.  
This simplification    
  greatly reduces the computational cost.  
A variant of the fast marching algorithms,  
  which attempts to retain 
  their computational advantages 
  over other level set methods 
  but allow for non-monotonically evolving fronts \cite{doi:10.1137/15M1017302} 
  was also developed 
  to give the flexibility 
  to account for non-uniform 
  interaction around the fronts. 
  
However, these models would not be idea for describing the propagation of fronts 
  driven by linear radiation emitted from a spatially finite source.  
Examples of scenarios such models 
  could describe is the propagation of ionisation fronts around radiative astrophysical objects such as stars 
  \citep[see e.g.][]{Goldsworthy1961RSPTA}
  or photochemical reactions 
  in a medium \citep[e.g.][]{Ducharme1990PhysRevB}
  irradiated by a light source.

In this paper,  
  we present a new model 
  of front propagation 
  which was specifically 
  developed to describe hydrogen ionisation 
  around radiative astrophysical objects, 
  but which can be applied to any scenario
  in which a source drives a front 
  radially outwards.   
This model divides the space 
  around the object 
  into a polar grid of cells, 
  each of which contains a number of particles and, 
  over a series of time steps, 
  models the progression 
  of the front as it transmits through these cells.  
The precision of the model 
  ultimately depends upon the sizes of the cells   
  and time steps which, 
  when the model is applied numerically, 
  will be limited only by the computational resources available.  
A similar method  
  was developed in \cite{Altay:2015lua} 
  for the study of astrophysical ionisation 
  in media of homogeneous densities, 
  but their study differed 
  in its aims and applications 
  from ours.

In Sec~\ref{sec:model}, 
  we introduce the model and describe its workings in a general scenario.  
In Sec.~\ref{sec:star}, 
  we apply the model to the case of 
  a Population-III star during the cosmological Epoch of Reionisation (EoR) 
  and compare the results we obtain 
  to those from a common analytic approximation, 
  showing that the results 
  are not only consistent 
  but provide us 
  with richer information 
  than those pre-existing methods.  
Finally, in Sec.~\ref{sec:disc}, 
  we highlight the capabilities of this model 
  and its potential for modelling front propagation 
  in complex media 
  that researchers encounter 
  in a variety of disciplines
  both within and beyond physics.

\section{The Radiative Front Model}
\label{sec:model}

Our model begins with a very generic assumption:  
  radiation is being emitted from a point source 
  and travelling radially away from that source.  
As such, the model would be applicable 
  to any system 
  which features a source transmitting radiation into a surrounding medium which could be expected to illicit an alteration in that medium with minimal scattering.  

The space around the source 
  is divided into cells arranged into a polar grid,  
  with the source at the origin.  
Taking the example of such a grid in 3D space, the volume of each cell is given by
\begin{equation}
    V_{i,\theta,\phi}=\frac{4}{3}\frac{\pi}{N_{\theta}N_{\phi}}
    \left[\;\! 
    (r_{i+1,\theta,\phi})^{3}-(r_{i,\theta,\phi})^{3}
    \;\! \right]
    \ ,
    \label{eq:cellvolume}
\end{equation}
   where $N_{\theta}$ is the number of cells 
      defined in the $\theta$ dimension, 
    $N_{\phi}$ is the number of cells 
    defined in the $\phi$ dimension, 
   and $r_{i,\theta,\phi}$ is the inner radius 
      of the cell in question.  
The medium under consideration is divided 
  into particles
  distributed between the cells 
    in accordance with the nature of the system being modelled.  
The number of particles in a given cell 
  with coordinates $\left(t,i,\theta,\phi\right)$ 
  that are in the initial state 
  will be labelled $n_{\mathrm{I};t,i,\theta,\phi}$,  
  and the number in the affected state 
  will be labelled $n_{\mathrm{II};t,i,\theta,\phi}$.  

Once the initial grid has been established, 
  the radiating object is treated as being activated. 
Assuming the speed of sound of its radiation 
 is a known function, which we label as $v$, 
 the maximum distance 
 that radiation could have travelled radially 
 from the source is computed at each time step 
 as $v(t-1)\Delta t$, where $\Delta t$ is the length of each time step, such that $(t-1)\Delta t$ is the time that has elapsed 
 by the beginning of that time step.  
Every cell that is within this distance 
 is then treated as being 
 in contact with the radiation. 
   
The medium in question has 
  an absorbativity parameter associated with it.  
In the case of a physical or chemical reaction 
  this will be a function, 
  $f$, of the particle cross-section 
  of the substance and the particle number in a given cell. 
We can therefore calculate the probability 
   of each unit of radiation 
   affecting a given unit of medium in a given cell as
\begin{equation}
    \Pi_{t,i,\theta,\phi}=1-e^{f_{t,i,\theta,\phi}} \ ,
\end{equation}
   such that the number of affected medium particles 
   per time step will be given by 
\begin{equation}
    \Gamma_{t,i,\theta,\phi}
    =J_{t,i,\theta,\phi}\;\! \Pi_{t,i,\theta,\phi}  \ , 
\end{equation}
    where $J_{t,i,\theta,\phi}$ is the flux of radiation entering that cell.

In order to calculate $J_{t,i,\theta,\phi}$, 
   we will need to not only account for the angular distribution of radiation but also for the absorption 
   in cells radially preceding the cell in question.  
We define the radial absorption function as
\begin{equation}
    s_{t,i,\theta,\phi}=\sum_{j=1}^{i-1}f_{t,j,\theta,\phi}
\end{equation}
such that $s_{t,i,\theta,\phi}$ gives the fraction of radiation that reaches the $i$th radial cell in a given angular bin.  Thus, the flux of radiation entering the cell will be given by
\begin{equation}
J_{t,i,\theta,\phi}=J_{t,1,\theta,\phi}
 \ e^{s_{t,i,\theta,\phi}} 
 \ .
\end{equation}

Having calculated the influx of radiation into a given cell at a given time step, we  recalculate the number of affected and unaffected particles as 
\begin{align}
    n_{\mathrm{I};t,i,\theta,\phi}&=n_{\mathrm{I};,t-1,i,\theta,\phi}-\Gamma_{t,i,\theta,\phi} \ ,\\
    n_{\mathrm{II};t,i,\theta,\phi}&=n_{\mathrm{II};,t-1,i,\theta,\phi}+\Gamma_{t,i,\theta,\phi} \ ,
\end{align}
in a manner that the overall particle count in a given cell 
  is preserved.  
 Any other effects which might affect the particle counts, 
   such as the stochastic processes 
   that can reverse the radiation's effects, 
   can be accounted for by adding additional terms into these equations.  
Having recalculated the particle numbers in each cell 
  at a given time step, 
  the absorption fraction can either be calculated again as a function of that cell's parameters at that time step 
  or more simply recalculated as
\begin{equation}
    f_{t+1,i,\theta,\phi}=f_{t,i,\theta,\phi} \ 
   \left(  \;\!
   \frac{n_{\mathrm{I};t,i,\theta,\phi}}{n_{\mathrm{I};0,i,\theta,\phi}} \;\! 
   \right)  \ ,
\end{equation}
   accounting for the alteration in the number of reactive particles in the medium at a given time step by taking their ratio with the initial number, 
   and the radial absorption function 
   would then be recalculated in kind.

\begin{figure}[h]
    \centering
    \includegraphics{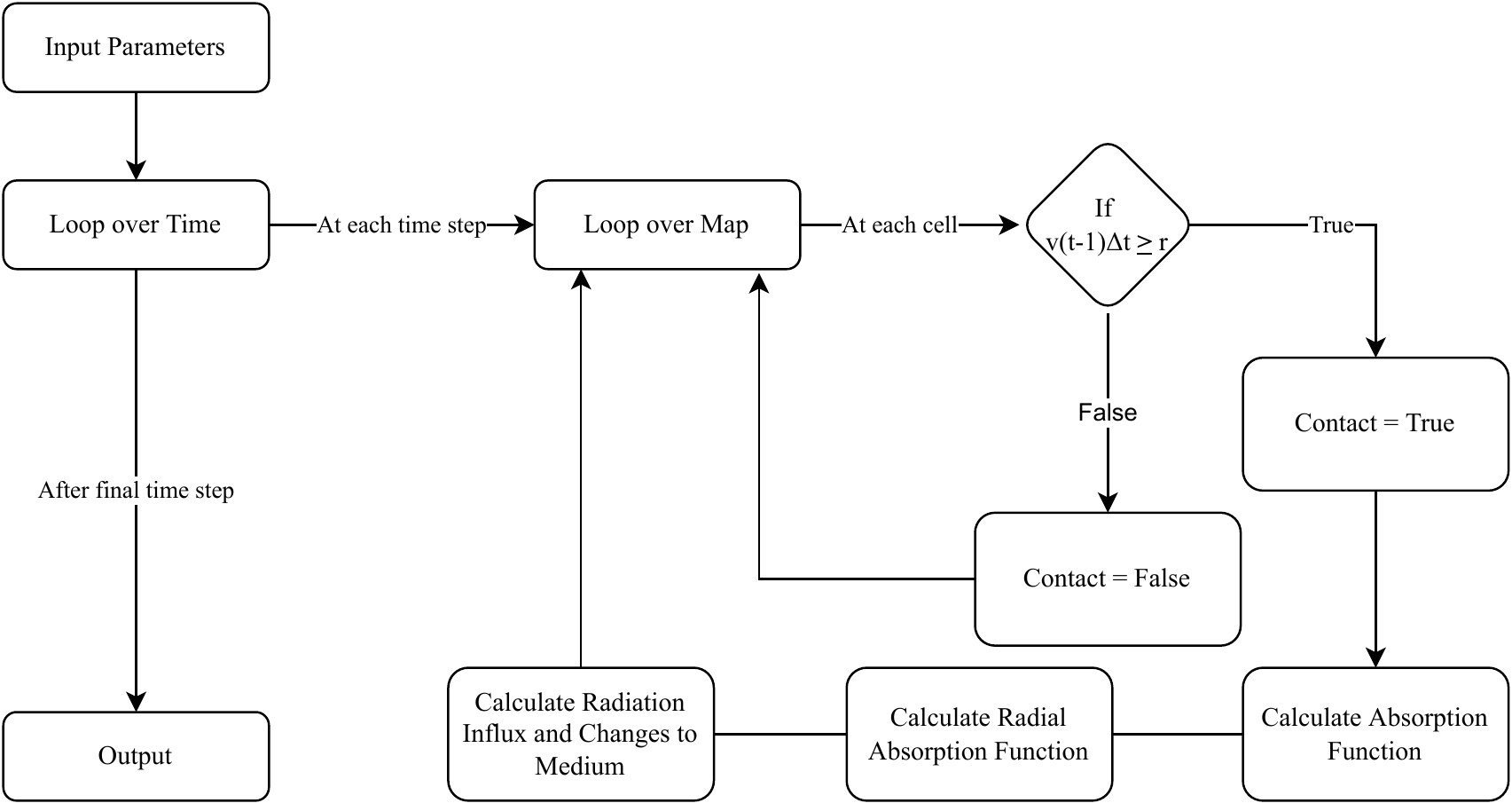}
    \caption{A flow chart showing the basic underlying mechanism of the radiative front model.  $r$ is the radial distance to the inner edge of the cell in question, $t$ is the time step coordinate, $\Delta t$ is the real time duration of each time step, and $v$ is the speed of the radiation causing the front, such that the cell in question can be considered to be in contact with the radiation in any timestep for which $v(t-1)\Delta t>r$.  In the case of a background which allows for non-trivial transport of the medium through which the front is propagating, transport terms will be included in the cell parameter calculations to represent this.  In the event that the radiation is travelling at less than the speed of sound in that medium, an additional series of statements would need to be added within the loop over the map to find which cells are within range of medium transport and recalculate those cell's parameters even when they are out of range of the radiation.}
    \label{fig:flow}
\end{figure}

In effect, at every time step and every cell, 
  we are calculating the amount of radiation reaching that cell 
  from the source, 
  taking into account absorption by preceding cells in the same angular bin, calculating the reaction rate in the cell due to that radiation, and preparing our map for a new set of calculations to take place at the next time step, 
  taking into account the ensuing changes in the medium. 
The mechanisms of this model is illustrated  
  in the flow chart in Fig.~\ref{fig:flow}.

\section{Example Application - Ionisation in Astrophysical Media}
\label{sec:star}
\subsection{Background}
\label{sec:starbackground}  

The earliest stars were composed almost entirely of hydrogen and helium,  and they are known as Population-III stars.  
\cite{Glover2005SSRv,Wise2011nha}.  
They were large and bright and emitted predominantly ultraviolet radiation. 
They were also short lived, with lifespans generally 
  not more than a few million years.  
They are believed to have first come into existence 
  during the transition 
  from what are called the cosmological dark ages, 
  during which the baryonic component of the Universe 
  was primarily composed of neutral hydrogen (HI), 
  to the EoR, during which that HI was predominantly converted into the ionised hydrogen (HII) that composes most of the baryonic matter of the Universe today
   \cite{Loeb:2000fc,Liu:2020krj}, beginning at a redshift\footnote{In cosmology, redshift is a measure of the expansion of the Universe; at a redshift of $z=15$, objects in the Universe will have been, on average, $1/15$th as far apart as they are now.} of approximately $z=15$ \cite{Koopmans:2015sua}.

An understanding of the ionisation dynamics 
  around individual objects as Population-III stars 
  is essential for high precision studies of the EoR, 
  in particular, those being carried out 
  by the 21cm surveys 
  \citep{DiMatteo:2004jha,Pritchard2012RPPhy,refId0} 
  though there is no precise model 
  able to describe in detail 
  the evolution, spatial extent and morphological structures  
  of the fronts these sources create  
  which determine the tomographic features 
  to be observed in the 21cm spectral lines 
  \cite{Mirocha:2017xxz}.

When ionising light travels through a medium containing neutral atoms, 
  each photon will have a probability of ionising an atom of 
\begin{equation}
    \Pi=1-e^{-\tau} \ ,
\end{equation}
where 
\begin{equation}
    \tau=\sigma d
\end{equation}
is the optical depth of the path traversed for interaction cross-section $\sigma$ and the column density of the path $d=\rho_{\mathrm{HI}} l$ 
   for HI number density $\rho_{\mathrm{HI}}$ and distance $l$.  
In the case of neutral hydrogen (HI) 
  being converted to ionised hydrogen (HII), 
  $\sigma=6.3\times 10^{-18}~\mathrm{cm^{-2}}$.

\begin{equation}
    \Gamma_{\mathrm{B}}=n_{\mathrm{e}}n_{\mathrm{p}}\alpha~,
    \label{eq:gammaB}
\end{equation}
for electron number density $n_{\mathrm{e}}$, proton number density $n_{\mathrm{p}}$, and B-recombination coefficient \cite{Osterbrock2006,1959MNRAS.119...90S,1964MNRAS.127..145P}
\begin{equation}
    \alpha\approx \frac{2.6\times 10^{-13}}{(T/10^{4}{\rm K})^{4/5}}
\end{equation}
with units of volume per unit time, where $T$ is the average temperature of the medium.

Due to B-recombinations, there will always be a theoretical limit to how large a photoionised bubble can be for a given radiation source.  That limit, given under the assumption that the system remains radiating and unchanging except for the interaction of the radiation and the medium for all of time, is given by the Str\"{o}mgren radius \cite{1939ApJ....89..526S}:
\begin{equation}
    r_{\mathrm{S}}=\left(\frac{3\phi}{4\pi n_{\mathrm{e}}n_{\mathrm{p}}\alpha}\right)^{\frac{1}{3}}~,
    \label{eq:RS}
\end{equation}
where $\phi$ is the total flux of photoionising particles being emitted.  This can be trivially rearranged to see that it is defined as the radius of a sphere around a radiative object at which the flux of radiation is equal to the number of B-recombinations; assuming that each photon can ionise only one particle, this will of course be the limiting radius of an ionisation bubble.  In the event that the ionising particles are able to ionise more than one atom each, this could be corrected by simply multiplying $\phi$ by the number of atoms each particle could ionise.

Of course, Eq.~\eqref{eq:RS} is not an entirely accurate model even for the idealised assumptions we are making; since the flux of particles will decrease with distance from the source, the number density of electrons and protons will also decrease, such that $r_{\mathrm{S}}$ should more realistically be phrased as an integral over spacetime with variable number densities.  Since Eq.~\eqref{eq:RS} would generally be calculated under the assumption that within the ionised region the number densities of ionised products equals the initial number density of neutral hydrogen and, if we were to proceed with such a more detailed analogue of the Str\"{o}mgren radius, we would expect that they would decrease with distance, we would find that such a calculation would yield a slightly higher result.  Therefore, we would expect the radius calculated with Eq.~\eqref{eq:RS} to actually be slightly smaller than the maximum radius reached by the ionisation front of a star.  Effectively, Eq.~\eqref{eq:RS} assumes a hard cutoff at which complete ionisation gives way to complete neutrality, while failing to take into account the existence of an intermediating front at all.

\subsection{Implementation}

We wish to model the physics described in Sec.~\ref{sec:starbackground} with the model we described in Sec.~\ref{sec:model}.  Specifically, we wish to model the evolution of the ionisation front around a single Population-III star in a neutral background medium during the EoR by treating the star as a source of radiation at the origin of a polar coordinate system and studying the effects of the radiation on the matter in the cell's that compose the spherical polar map.  The speed of the radiation is the speed of light such that cells are considered to be in contact with the source when $r_{i,\theta,\phi}\leq c(t-1)\Delta t$ where $c$ is the speed of light and we are assuming a negligible difference between the speed of light in a vacuum and the speed of sound of the light in the medium.

The absorption function of a given cell is given by its optical depth
\begin{equation}
    \tau_{t,i,\theta,\phi}=\sigma \;\!  \rho_{\mathrm{HI};t,i,\theta,\phi}l~,
\end{equation}
for cross-section $\sigma$ and cell depth $l=r_{i+1}-r_{i}$.  From the definition of $\tau$, we have that $e^{-\tau}$ is the fraction of radiation that transmits through the medium without scattering.  

The radial absorption function is then given by a sum over the optical depths of the preceding cells in the same angular bin,
\begin{equation}
s_{t,i,\theta,\phi}=\sum_{j=1}^{i-1}\;\! 
 \tau_{t,j,\theta,\phi} \ .
\end{equation}

In order to calculate the flux 
  of radiation reaching each cell 
  at each time step, 
we must not only account for the division of the star's overall flux throughout the angular components of the map, but also for the absorption that takes place before they reach a given cell.  Thus, the radiation intensity at a given cell will be a function of the overall stellar luminosity\footnote{Luminosity is defined as being the overall radiation flux emitted by a light emitting source.} $L$, the number of angular components $N_{\theta}$ and $N_{\phi}$, and the radial optical depth $S_{t,i,\theta,\phi}$.

Thus, at a given cell which we label cell $i$, we have that the ionising radiation intensity is given by
\begin{equation}
    J_{t,i,\theta,\phi}=\frac{ Le^{-s_{t,i,\theta,\phi}}}{N_{\theta}N_{\phi}}\Delta t~.
\end{equation}
The ionisation rate within that cell is then given by
\begin{equation}
\Gamma_{t,i,\theta,\phi}=\frac{JN}{h\nu}(1-e^{-\tau_{t,i,\theta,\phi}}) \ ,
\label{eq:gammai}
\end{equation}
where $N$ is the number of atoms each photon is capable of ionising, given by $N=\lfloor h\nu/E_{\mathrm{i}}\rfloor$ where $E_{\mathrm{i}}$ is the energy needed to ionise a ground state atom.  In the event that multiple types of radiation are emitted, we can generalise Eq.~\eqref{eq:gammai} to become a sum over the different wavelengths emitted or, in the event of a continuous distribution of radiation wavelengths, this could be further generalised to
\begin{equation}
\Gamma_{t,i,\theta,\phi}
 =\int {\rm d}j 
 \ \frac{J_{j}N_{j}}{h\nu_{j}}
 \;\! \left( 
  1-e^{-\tau_{j,t,i,\theta,\phi}}
  \right) \ ,
\label{eq:gammai3}
\end{equation}
where $N_{j}=\lfloor h\nu_{j}/E_{\mathrm{i}}\rfloor$.  However, for this simplified study we content ourselves to treat the star's radiation as being purely composed of $13.6\mathrm{eV}$ ultraviolet light; given the heavily ultraviolet spectrum of a Population-III star, this is not a particularly inaccurate treatment and will allow accurate results to a level of precision that is satisfactory for the present study.

Given that HI has only one electron, we treat $n_{\mathrm{e}}=n_{\mathrm{p}}=n_{\mathrm{HII}}$, such that Eq.~\eqref{eq:gammaB} can be implemented as
\begin{equation}
    \Gamma_{\mathrm{B};t,i,\theta,\phi}= {(n_{\mathrm{HII};t,i,\theta,\phi})}^{2}
    \;\! \alpha_{t,i,\theta,\phi} 
    \;\! V_{i,\theta,\phi}
    \;\! \Delta t \ .
\end{equation}

As a first approximation, 
  we treat the temperature of all cells as being constant, 
  at $10^{4}\mathrm{K}$.  
This is an appropriate temperature 
  of such ionised media\footnote{The temperature will be heavily dependent on the chemical composition of the medium; since this paper concerns itself with demonstrating the validity of the proposed method rather than a precise study of astrophysical ionisation, we content ourselves to use a constant temperature of $T=10^{4}~\mathrm{K}$ in both our analytic calculations with Eq.~\eqref{eq:RS} and our implementation of the radiative front model.}.  
Furthermore, we treat the astrophysical medium as being entirely composed of hydrogen, such that at all times we will have $n_{\mathrm{e}}=n_{\mathrm{p}}$ and 
$N=\lfloor\;\! h\nu/13.6\;\!{\rm eV} \;\!\rfloor$.  
In a detailed study relevant to 21cm surveys, we would want to use a more detailed analysis with variable temperatures and a medium containing helium, but for the purposes of demonstrating the radiative front model and its application to the evolution of photoionised fronts, these approximations are valid, particularly when one bears in mind that the Str\"{o}mgren radius given in Eq.~\eqref{eq:RS} assumes a purely hydrogen background and is itself a function of temperature, which we can also set to a constant $10^{4}~\mathrm{K}$.  

Thus, at each time step we are altering each cell's HI and HII count by 
\begin{align}
    n_{\mathrm{HI};t,i,\theta,\phi}&=n_{\mathrm{HI};t-1,i,\theta,\phi}-\Gamma_{t,i,\theta,\phi}+\Gamma_{\mathrm{B};t,i,\theta,\phi}~,\\
    n_{\mathrm{HII};t,i,\theta,\phi}&=n_{\mathrm{HII};t-1,i,\theta,\phi}+\Gamma_{t,i,\theta,\phi}-\Gamma_{\mathrm{B};t,i,\theta,\phi}~,
\end{align}
in a manner that naturally preserves the overall number of particles per cell, before recalculating the optical depths as
\begin{align}
    \tau_{t+1,i,\theta,\phi}&=\sigma \rho_{\mathrm{HI};t,i,\theta,\phi}l~,\\
    s_{t+1,i,\theta,\phi}&=\sum_{j=1}^{i-1}\tau_{t+1,j,\theta,\phi}~,
\end{align}
in preparation for the next time step.

\subsection{Application}

We demonstrate the model by taking the example of a $100~\mathrm{M_{\odot}}$ Population-III star in a homogeneous background composed entirely of HI with a number density of $\rho_{\mathrm{HI}}=5\times 10^{-6}\;\!(1+z)^{3}\mathrm{~cm^{-3}}$ 
 \cite{Crighton:2015pza}, where the bracketed term accounts for the reduction in density with the expansion of the Universe. Such a star would have a lifespan of approximately $2.7~\mathrm{Myr}$ and would emit 
$2.4\times 10^{51} \mathrm{~eV~s}^{-1}$ 
of predominantly ultraviolet radiation \cite{Wise2011nha}, 
which for the sake of simplicity we treat as being purely $13.6~\mathrm{eV}$ light.  With these parameters, we generated homogeneous density maps with densities equal to the average densities at $z=15$, $z=10$, and $z=5$.  

\begin{figure}[h]
    \centering
    \includegraphics[width=.49\linewidth]{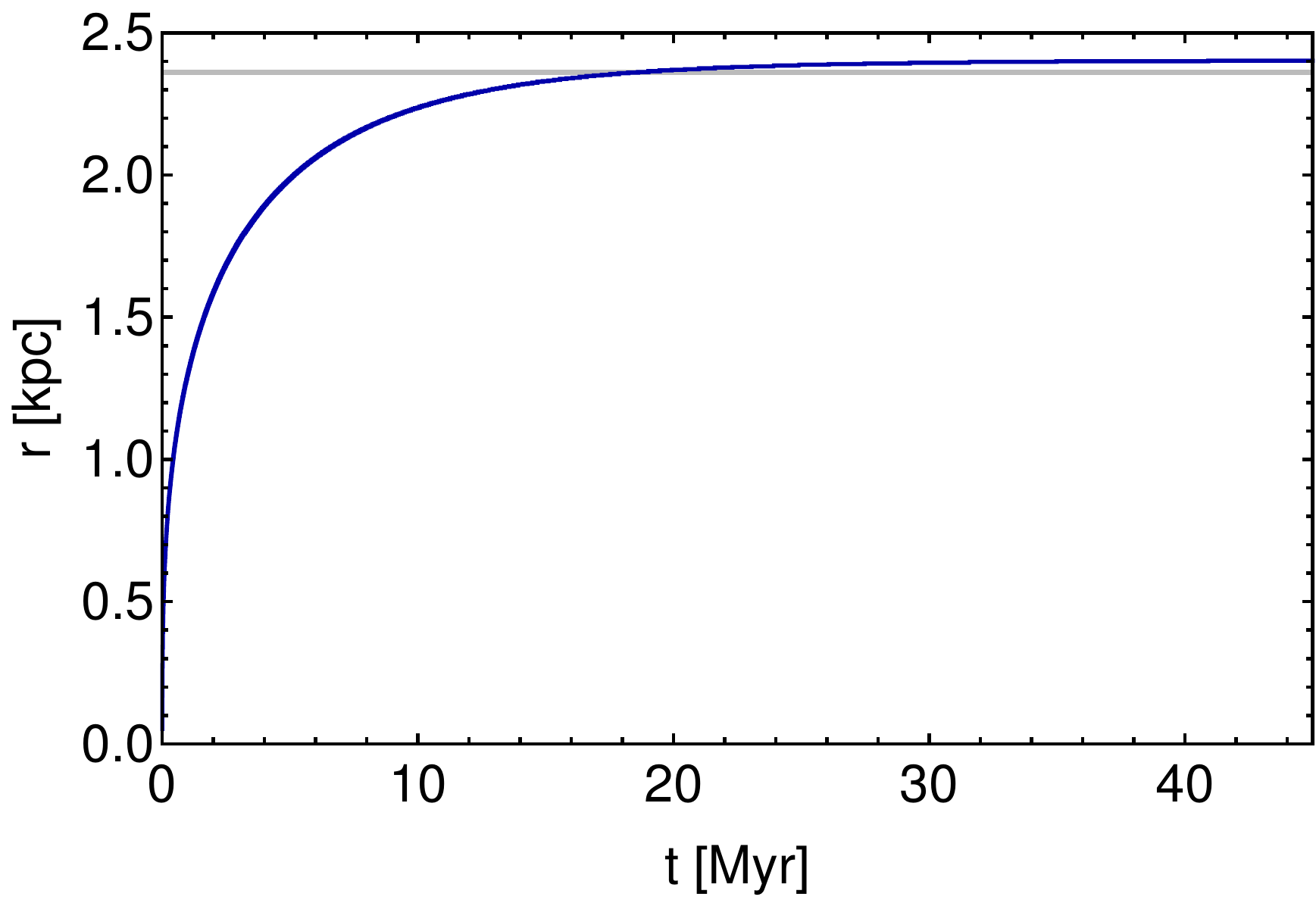}
    \includegraphics[width=.49\linewidth]{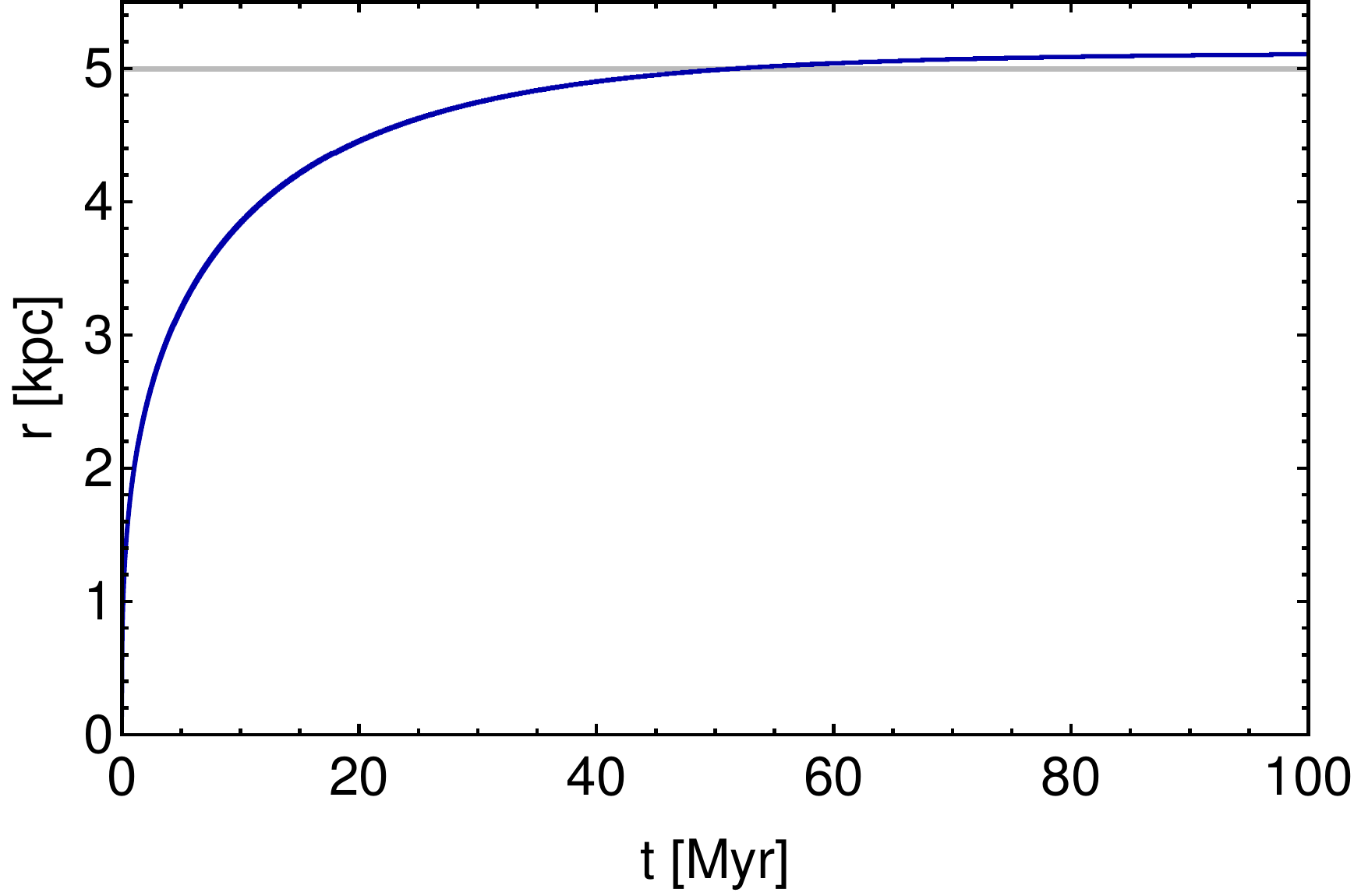}
    \includegraphics[width=.49\linewidth]{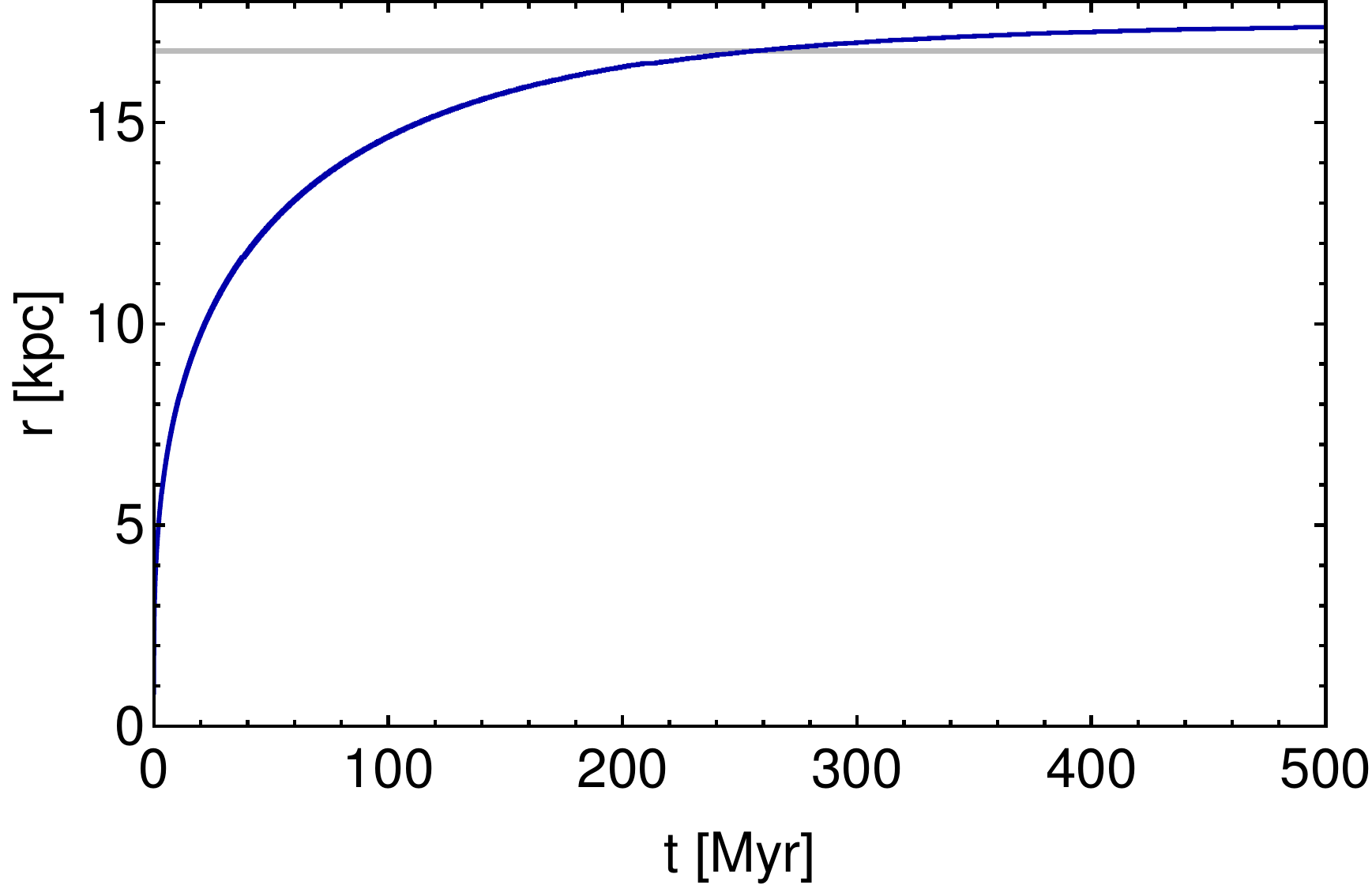}
    \caption{The maximum extent of the ionisation front, defined as being the radius of the cell at which $1-n_{\mathrm{HI},t,i,\theta,\phi}/n_{\mathrm{HI,0},t,i,\theta,\phi}>0$ and $1-n_{\mathrm{HI},t,i+1,\theta,\phi}/n_{\mathrm{HI,0},t,i+1,\theta,\phi}=0$, over periods of time far greater than the realistic lifespan of a Population-III star assuming a constant density as given by the average densities at $z=15$ (top left panel), $z=10$ (top right panel), and $z=5$ (bottom panel), showing that, as expected, the ionisation fronts slightly exceed the Str\"{o}mgren radii calculated from Eq.~\eqref{eq:RS}, plotted alongside the model's results as grey lines.}
    \label{fig:zmax}
\end{figure}

Before using the model to obtain new results, we would like to confirm that it is consistent with existing ones.  
In order to do this, we allow the stars to exist for unrealistically long lifespans and compare the maximum radii of their ionised bubbles to the Str\"{o}mgren radii calculated from Eq.~\eqref{eq:RS}.  
In Fig.~\ref{fig:zmax}, we plot the maximum extent of the front against the Str\"{o}mgren radii calculated from Eq.~\eqref{eq:RS} at the same three redshifts, where the maximum extent of the front is defined in the algorithm as being the inner radius of the cells at which $1-n_{\mathrm{HI};t,i,\theta,\phi}/n_{\mathrm{HI};0,i,\theta,\phi}>0$ and $1-n_{\mathrm{HI};t,i+1,\theta,\phi}/n_{\mathrm{HI};0,i+1,\theta,\phi}=0$. In these cases, we adjusted the run time of the model to account for the respective times it took for the front to reach an equilibrium at the different redshifts, changing the length of each time step accordingly, as well as altering the radius of each cell.  In the $z=15$ case, we used cell radii of $2.5~\mathrm{pc}$ and time steps of $450~\mathrm{years}$, in the $z=10$ case we used a cell radius of $5.2~\mathrm{pc}$ and time steps of $1000~\mathrm{years}$, and in the $z=5$ case we used cell radii of $18~\mathrm{pc}$ and time steps of $5000~\mathrm{years}$.  Due to the isotropy of the map, we used $N_{\theta}=N_{\phi}=1$ in order to maximise computational efficiency.

We can see that the curves end up slightly exceeding the Str\"{o}mgren radius, as expected.  We also see that in higher densities, ionisation fronts tend to reach this equilibrium position sooner; in the case of $z=5$, for example, it took approximately $500~\mathrm{Myr}$ for the front to stop expanding, while in the case of $z=15$ it too approximately $50~\mathrm{Myr}$.  This is in keeping with the notion that, in higher density, regions, the higher rate of B-recombinations will bring about a total radiation absorption in a smaller time frame as well as a smaller volume.

Having confirmed that the model is consistent with expected results, we may use it to generate new results by studying the composition of the photoionised region and the front that borders it in more detail.   We therefore set the stars to exist for their realistic lifespan of $2.7~\mathrm{Myr}$ and ran the model with time steps of $2000~\mathrm{years}$ and cell radii of, in the case of $z=15$, $17.14~\mathrm{pc}$, in the case of $z=10$, $8.57~\mathrm{pc}$, and in the case of $z=5$, $5~\mathrm{pc}$, once again setting $N_{\theta}=N_{\phi}=1$ for efficiency since we are assuming a homogeneous cosmological background.  

\begin{figure}[h]
    \centering
    \includegraphics[width=.49\linewidth]{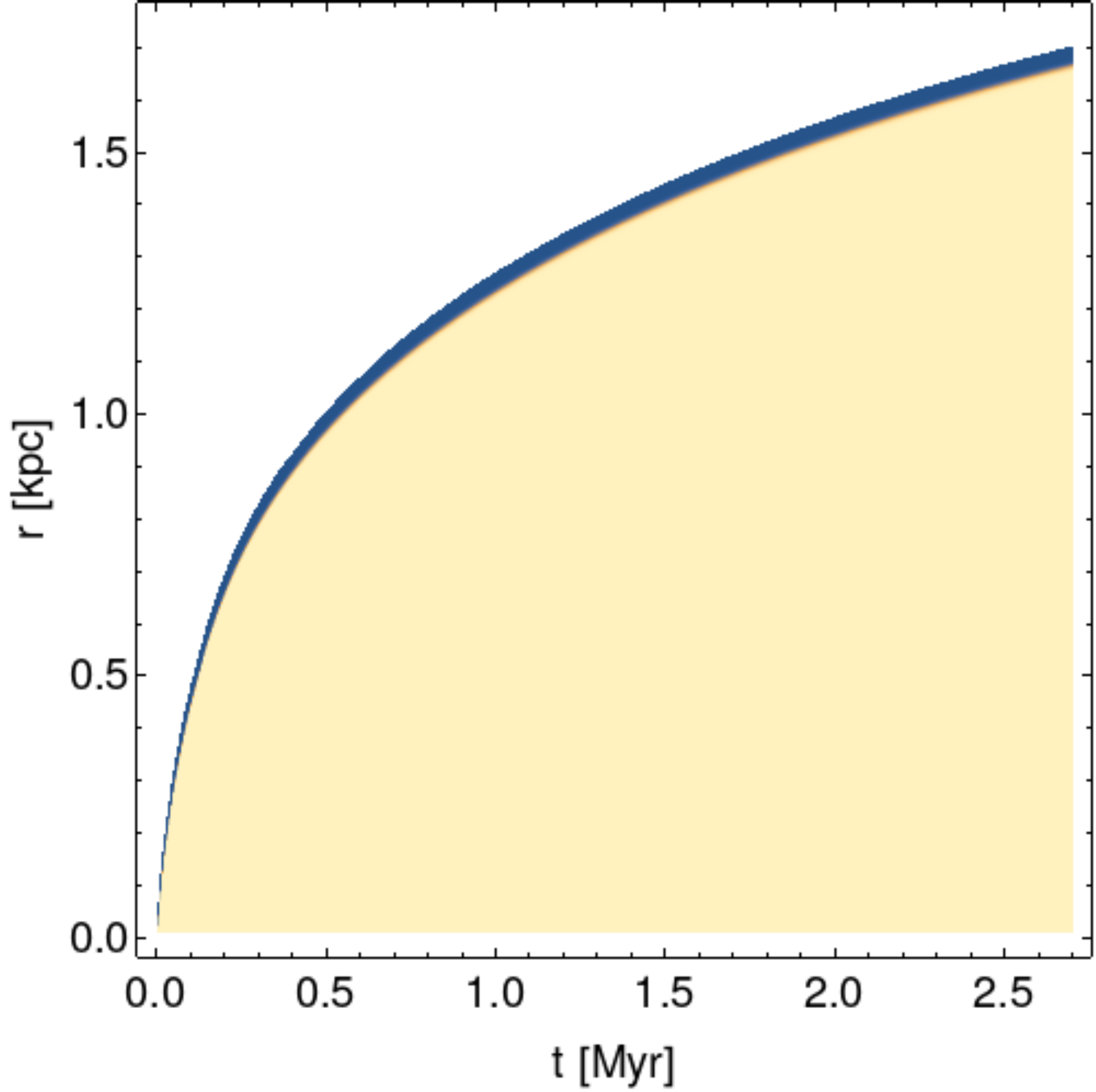}
    \includegraphics[width=.49\linewidth]{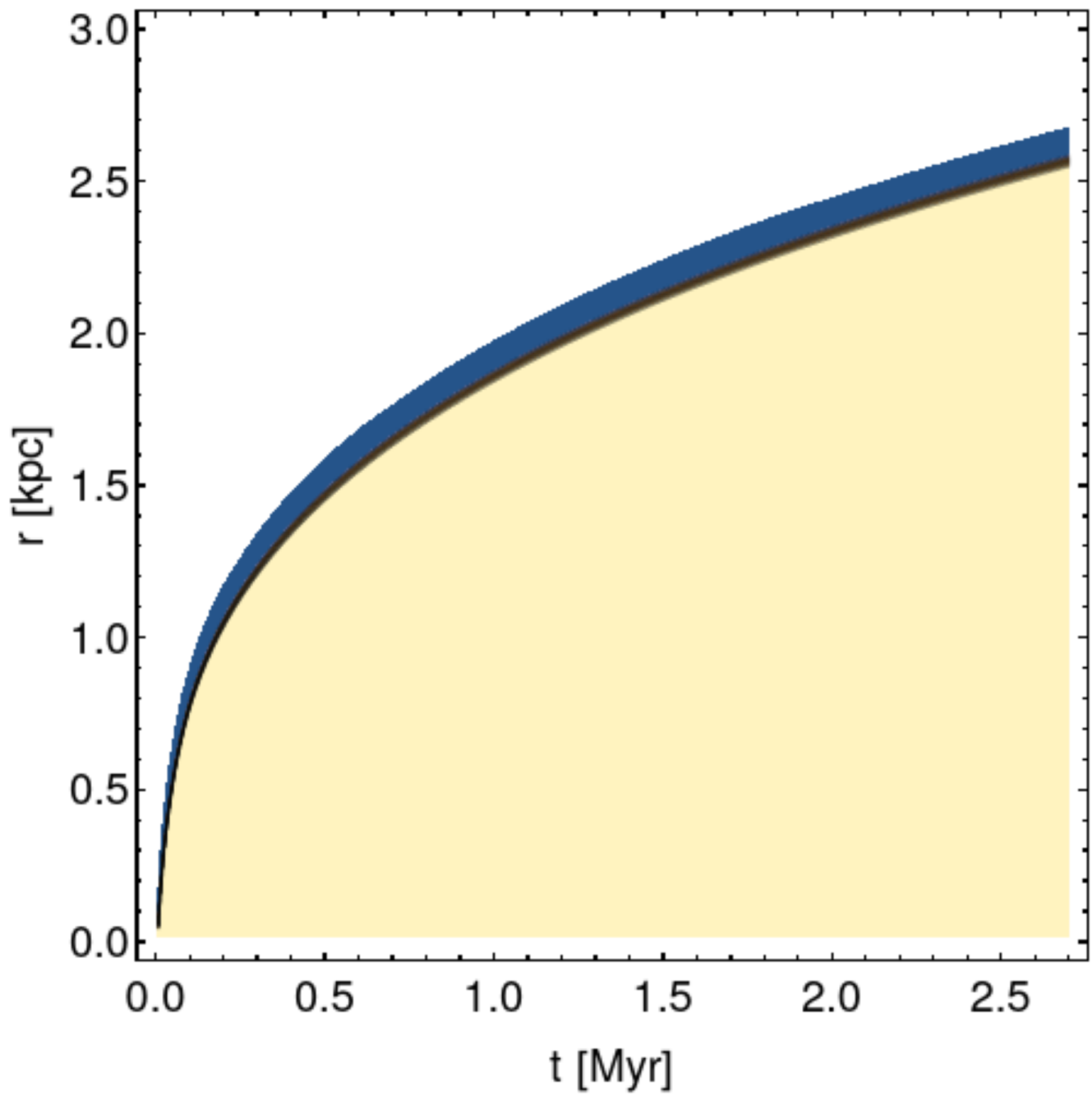}
    \includegraphics[width=.49\linewidth]{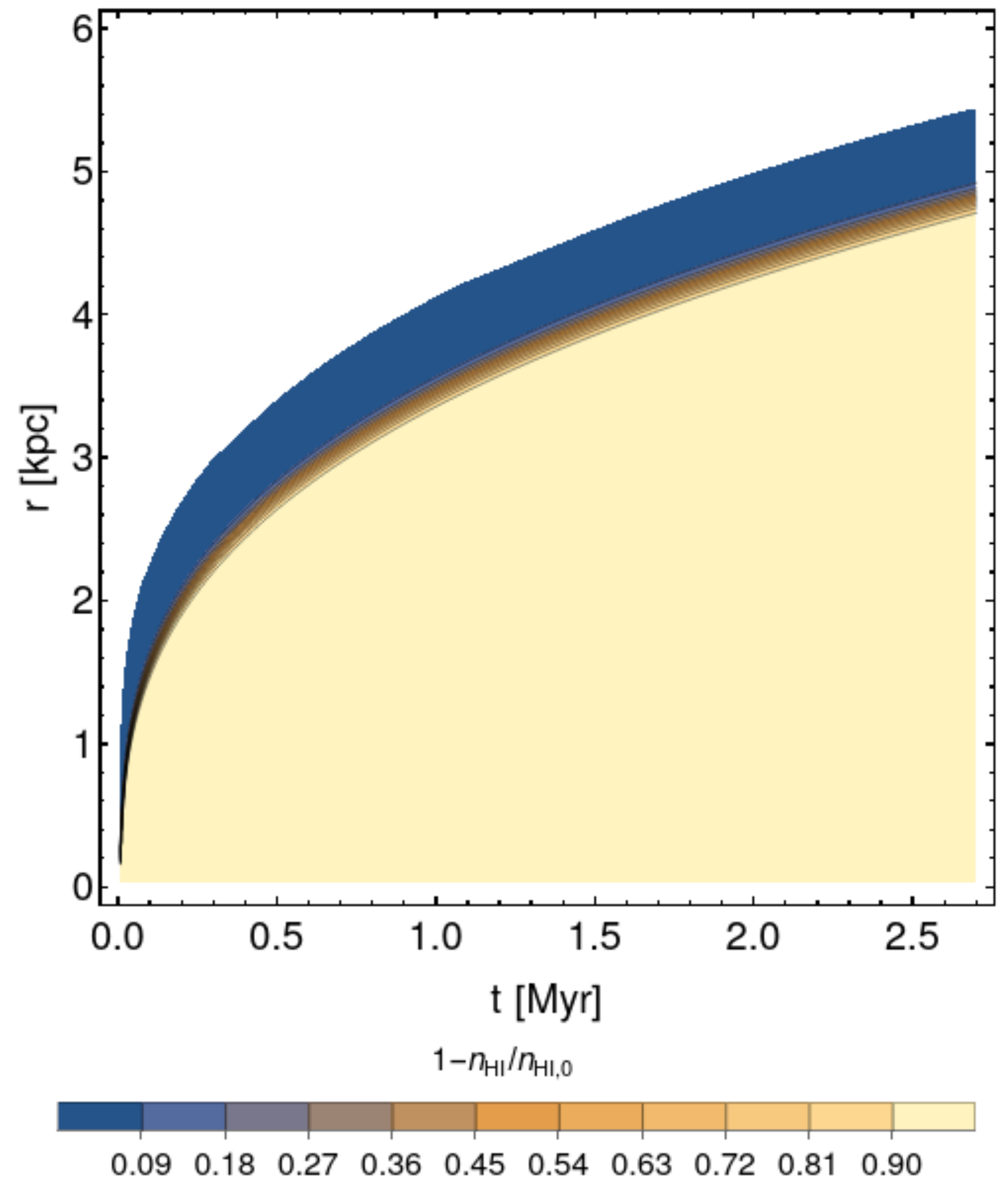}
    \caption{The evolution in time of the photoionisation front created by a Population-III star emitting $2.4\times 10^{51} \mathrm{~eV~s}^{-1}$ of purely $13.6~\mathrm{eV}$ ultraviolet light consistently from its creation to its destruction at an age of $2.7 \mathrm{Myr}$ at $z=15$ (top left (panel), $z=15$ (top right panel), and $z=5$ (bottom panel), under the approximation of a constant background density at each redshift.  The hydrogen to the outside of the front is almost entirely neutral, that inside it is almost entirely ionised with a small amount of constant recombination preventing full ionisation from ever occurring, and the region within the front is defined as being the region that has a reduced ionised to neutral hydrogen ratio $1-n_{\mathrm{p}}/n_{\mathrm{H}}$ of between $0$ and $1$.  As one would expect, the photoionised bubble evolves more quickly and to a larger size in eras with a lower matter density and the front at a given time is wider, as the lower optical depth of the medium allows a greater depth of space to be being ionised simultaneously.  Cells in which $n_{\mathrm{HI}}=n_{\mathrm{HI,0}}$ at a given time step are not plotted and therefore make up the region beyond the blue curve of the outer layer of the front.}
    \label{fig:contours}
\end{figure}

Running the model thus, we obtain the results shown in Fig.~\ref{fig:contours}.  These figures show the evolution of the angularly symmetric front with time, defined through the ratio $1-n_{\mathrm{HI}}/n_{\mathrm{HI,0}}=n_{\mathrm{HII}}/n_{\mathrm{HII,0}}$, the fraction of the medium which has been ionised.  As can clearly be seen, photoionised bubbles in higher density media as are presented by the higher redshift samples evolve more slowly and with a thinner front.  Indeed, not only do the higher density samples present smaller bubbles and fronts, but the ratio of the width of the front to the width of the photoionised region is much lower, indicating that the radiation is able to penetrate partially ionised cells much less effectively due to their higher count of HI particles.  We also find that the front during the earlier phases of expansion is much thinner than this and only becomes thicker with a more apparent gradient as the ionised bubble becomes larger, which is what one might intuitively expect as the radiation becomes more dispersed and so begins to ionise the media it is in contact with at a lower rate.

\begin{figure}[h]
    \centering
    \includegraphics[width=.49\linewidth]{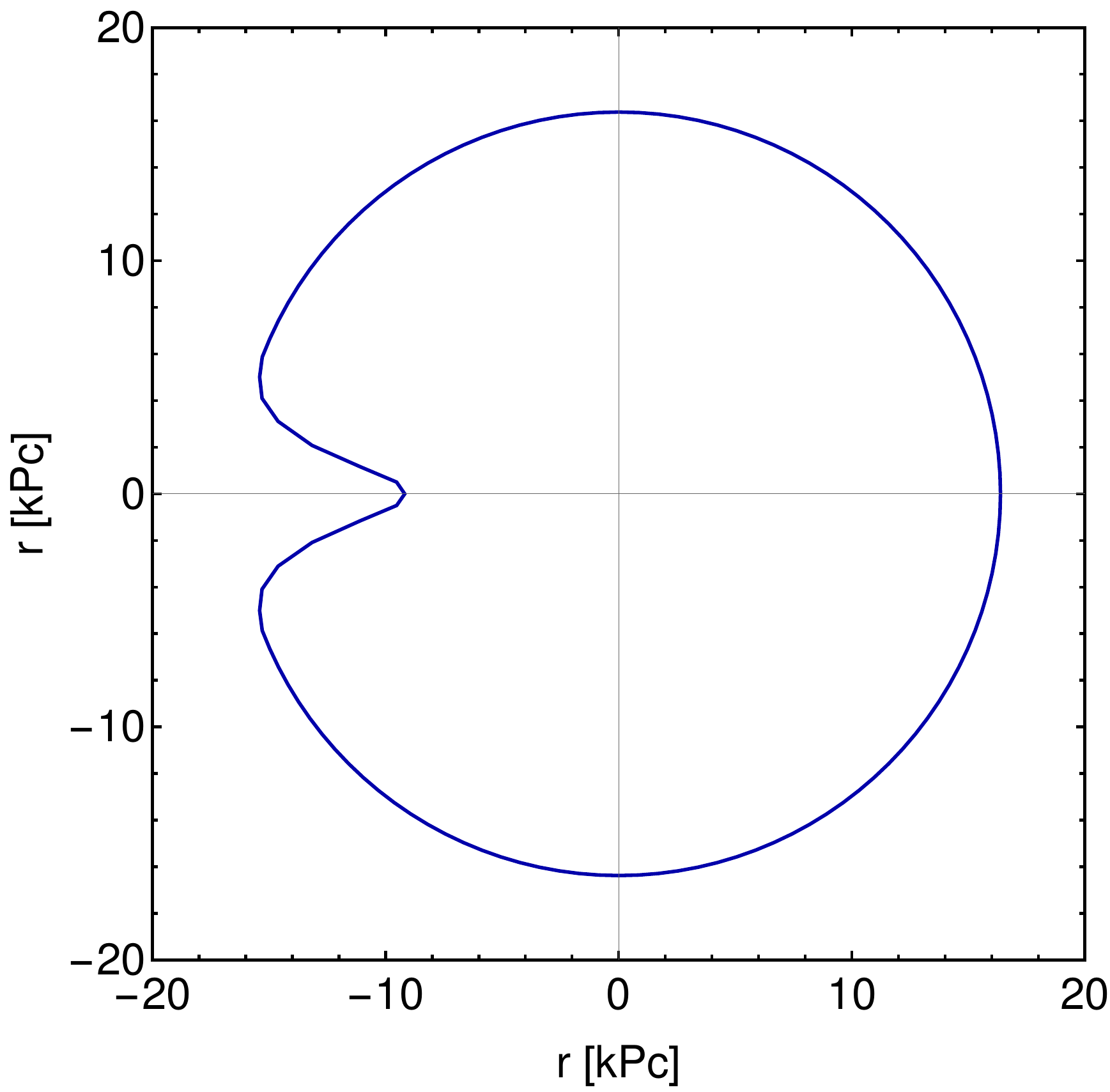}
    \caption{A slice in $\phi$ of the maximum extent of the ionisation front, defined as being the radius of the cell at which $1-n_{\mathrm{HI};t,i,\theta,\phi}/n_{\mathrm{HI,0};t,i,\theta,\phi}>0$ and $1-n_{\mathrm{HI};t,i+1,\theta,\phi}/n_{\mathrm{HI,0};t,i+1,\theta,\phi}=0$, after a period of time of $500~\mathrm{Myr}$ with a density contrast defined at $r\geq 9~\mathrm{kPc}$ around $\theta=3\pi/2$ and exponentially decreasing around that angle before vanishing at $\theta=0$ and $\theta=\pi$.}
    \label{fig:zmaxasymmetric}
\end{figure}

Taking another example, that of an inhomogeneous background density, once again with a supersonic ionisation front and the same stellar parameters, we define a density map in which the background is primarily defined as being the average density at $z=5$ with a region of overdensity at $\theta=3\pi/2$ which decreases exponentially around that centre until it vanishes at $\theta=0$ and $\theta=\pi$.  Allowing the star to live for the unrealistic timeframe examined for the results shown in Fig.~\ref{fig:zmax} and studying again the maximum extent of the front, we find that a slice in time and the homogeneous angular dimension gives a final ionisation front at $2.7~\mathrm{Myr}$ which varies with density, as shown in Fig.~\ref{fig:zmaxasymmetric}.  This serves to highlight the models ability to study supersonic radiative transport in inhomogeneous media, such as in a region of space containing nebulae and other sources of over- and underdensity.

\section{Discussion}
\label{sec:disc}
We have proposed a new method for modelling the evolution of fronts around radiative objects and have examined the example application of the ionisation bubble around a star in a neutral medium.  We have shown that this method agrees with existing methods for estimating basic parameters of the front while providing significantly more information than those methods, allowing us to precisely model its evolution in time as a function of all relevant physical parameters.  We propose that this method could find many uses in varied fields, from modelling the effects of ionising radiation as examined here, to modelling chemical photocatalysis, to studying more varied phenomenon such as the propagation of ant expeditions as they explore the territory around their nests.  We intend to continue the development of this model and use it for astrophysical applications, specifically using it to study ionising objects during the cosmological epoch of reionisation and how the information this model provides can be applied to upcoming studies of 21cm spectroscopy and we encourage members of other academic groups to apply this method with appropriate modifications and parameters to their own fields.

The current formulation of the Radiative Front Model relies upon a static medium through which the front is propagating; this works for a supersonic front, but would run into issues when describing subsonic fronts as they are much more likely to encounter alterations due to fluid transport caused by the change of state they herald.  Furthermore, the model assumes that no radiation is scattered to non-radiative angles.  

The subsonic front issue could be corrected for by introducing a set of medium transport equations into the model which allow the particle number to change between cells in a manner that preserves the overall particle number over the set of cells.  

The scattering of radiation could be accounted for in one of a number of ways; the simplest would be to incorporate a scalar field into the grid which affects each cell as a function of the conditions of its surrounding cells, effectively modelling the non-radiative scattering by taking an average of its expected effects and applying them directly to each cell.  Alternatively, the radiation could be modelled coming out of each cell in which non-radiative scattering is occurring directly before being mapped onto the initial polar coordinate system of cells; however, this method would be significantly more computationally expensive.

Another possible modification which could improve both the precision and computational efficiency of the model would involve dividing each cell into smaller subcells which could be selectively explored depending upon the circumstances; for example, regions well within or well outside the front could be treated as single units, while transitional front cells could be divided up to get a smoother description of the front as it evolves.  This could be performed by assigning regions that have been transitioned enough to be considered no longer a part of the front but a part of the altered region with a logical statement and then grouping all cells in contact with each other with that logical statement together as one larger cell, while taking all cells that are in contact with the radiation but are still considered to be a part of the front into smaller cells.  However, for the purposes of this paper which was intended to describe and demonstrate our new method, we restricted ourselves to the simpler example of using fixed cell sizes.

We leave it to a future paper to explore these potential extensions of the model in more detail and limit ourselves to the study of entirely radiative, supersonic fronts in this paper.

\bibliography{bibliography}

\end{document}